\documentclass{aastex63}
\journalinfo{}
\makeatletter
\let\frontmatter@title@above=\relax
\makeatother
\usepackage{}% for placeholder text

\begin{document}

\hfill \break
\hfill \break

\title{\Large Multiband Photometry Evolution in the First Weeks of SN 2023ixf, a possible II-L Subtype Supernova}

\author[0000-0003-1276-5812]{Bianciardi G.}
\affiliation{University of Siena-retired (Italy)}
\affiliation{UAI Remote Telescope (Italy) \\
}
\author[0000-0002-2449-1591]{Ciccarelli A.M.}
\affiliation{Stazione Astronomica dei Marsi (Italy) \\
}
\author[0000-0002-2412-1558]{Conzo G.}
\affiliation{UAI Remote Telescope (Italy) \\
}
\affiliation{MicroObservatory Robotic Telescope Network (USA) \\
}
\affiliation{Gruppo Astrofili Palidoro (Italy) \\
}
\author[0009-0005-8903-3312]{D'Angelo M.}
\affiliation{Gruppo Astrofili Palidoro (Italy) \\
}
\author{Ghia S.}
\affiliation{UAI Remote Telescope (Italy) \\
}
\author[0000-0002-5600-0953]{Moriconi M.}
\affiliation{Gruppo Astrofili Palidoro (Italy) \\
}
\author[0009-0005-0862-3215]{Orbanić Z.}
\affiliation{Orbanić Explorer Observatory (Croatia) \\
}
\author[0009-0006-0421-8957]{Ruocco N.}
\affiliation{Osservatorio Astronomico Nastro Verde (Italy) \\
}
\author[0009-0004-8987-2127]{Sharp I.}
\affiliation{Ham Observatory (United Kingdom)}
\affiliation{PixelSkies Remote Telescope (Spain)\\
}
\author[0009-0002-0636-326X]{Uhlár M.}
\affiliation{Observatory and Planetarium Prague (Czech Republic)}
\affiliation{Czech Astronomical Society (Czech Republic)\\
}
\author[0000-0003-2060-4912]{Walter F.}
\affiliation{Observatory and Planetarium Prague (Czech Republic)}
\affiliation{Czech Astronomical Society (Czech Republic)\\
}

\correspondingauthor{Bianciardi G.}
\email{giorgio.bianciardi@unisi.it}



\begin{abstract}

Multiband photometric observations and their evaluation to instrumental magnitudes were performed using standard Johnson-Cousins filters (B, V, Rc) as well r and g Sloan filters, and not standard ones (R, G, B, and Clear filters). These were recorded from 9 observatories and from the MicroObservatory Robotic Telescope Network. The results describe the rapid ascent towards the maximum (2.5 magnitudes about in five days in the B filter) and the slow decrease after the maximum (0.0425 ± 0.02 magnitudes/day in the B filter).  The results highlight the strong variation of the B-V colour indices during the first 50 days (from -0.20 ± 0.02 to +0.85 ± 0.02) and V-R (from 0 ± 0.01 to +0.50 ± 0.01) after the explosion, presumably corresponding to the cooling of the stellar photosphere.  At 50 days after the explosion the magnitude decrease from the maximum was observed to continue where it faded by 2.5 magnitudes (B filter), thus we propose SN 2023ixf is a Type II, subtype L, supernova (SNe).

\end{abstract}


%% and \cvolta che apro il docuitet commands to identify citations.  The citations are

\hfill \break

\section{Introduction} \label{sec:intro}
   Supernovae (SNe) can be classified as belonging to Type I or Type II depending, respectively, on the absence or presence of hydrogen in their spectra. The first studies on Type II SNe identified a value of absolute magnitude $M_B = -16.45$ (sigma = 0.78) reached at the maximum of their brightness \citep{1979A&A....72..287B}.
   For Type II SNe, we know that, since the work of Barbon et al 1979, two principal further divisions of Type II SNe can be made, based solely on the shape of the post-maximum light curves. These are the II-plateau (P), about 2/3 of the Type II SNe known, and II-linear (L) types. The former displays an extended period (typically 100d) during which the (VRI-band) brightness remains constant, usually to within $ \lesssim 0.5mag$. Conversely, type II-L SNe decline linearly at a rate of a few hundredths of a magnitude per day in the same time period. On May 19, 2023, a Type II supernova \citep{2013ApJ...777...79M} was observed by Koichi Itagaki \citep{2023TNSTR1158....1I} inside the galaxy M101, measuring an apparent magnitude of 10.8 in the visible spectrum. The galaxy which resides in the constellation of Ursa Major, is 6.19 Mpc from Earth \citep{2013ApJ...777...79M}. Therefore, this is the closest Type II SNe observed in decades. Its proximity allows detailed observations of the supernova, called SN 2023ixf, even by means of amateur telescopes. We have focused our study of Supernova SN 2023ixf on multiband photometric observations of the star for the first 50 days since the discovery of the supernova event, presenting observations started on May 19, 2023, the same day of its discovery. Different sets of filters were used - standard photometric and tri-colour sets, according to the different observers located in different countries of the European continent who ranged from the UK to the south of Italy. This resulted in a wider temporal coverage because the chances of a clear sky were increased. Some low-resolution spectra were also performed during the days of observation.\\

%ref ready for the bibliography %
%\pagebreak

\hfill \break

\section{Observations and data analysis} \label{sec:floats}

%\subsection{Photometric investigation}

\begin{figure}[ht]
\begin{center}
\includegraphics[width=14cm]{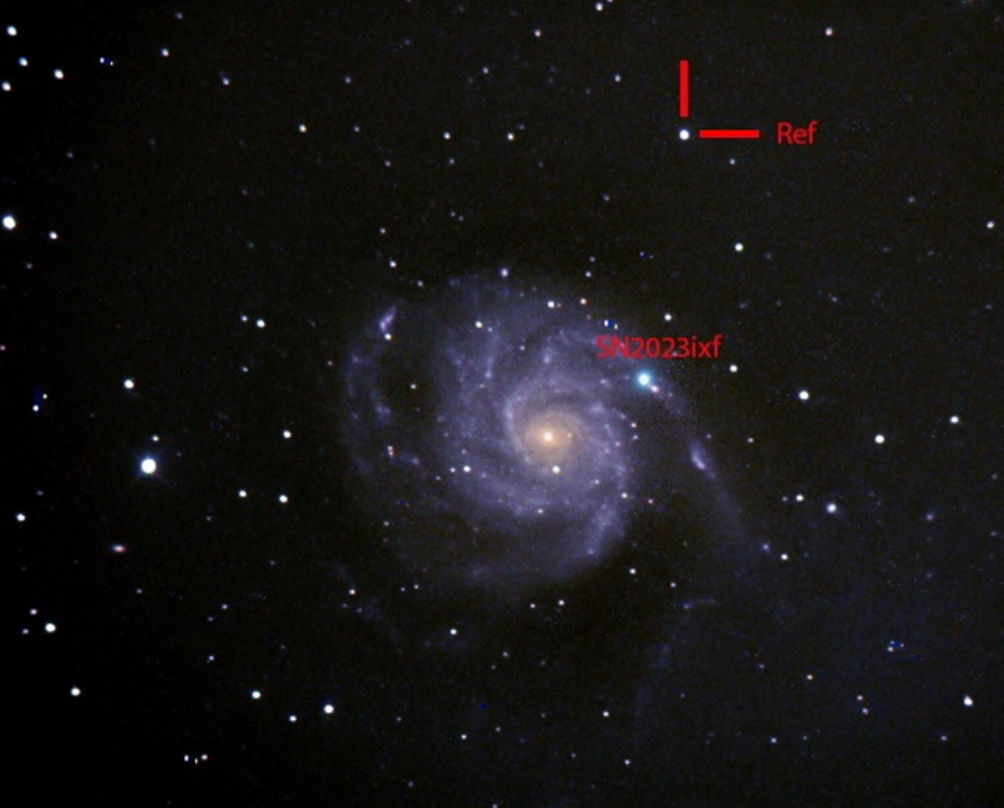}
\caption{SN 2023ixf occurred in a spiral arm of the Pinwheel Galaxy. This wide-field image was made by one of us on the night of 2023/05/28. The characteristic cyan colour of the supernova near the maximum is well-defined. Reference star is indicated. Image credit: Giuseppe Conzo, UAI Remote Telescope.}
\label{fig:fig1}
\end{center}
\end{figure}

\begin{figure}[ht!]
\begin{center}
\includegraphics[width=14cm]{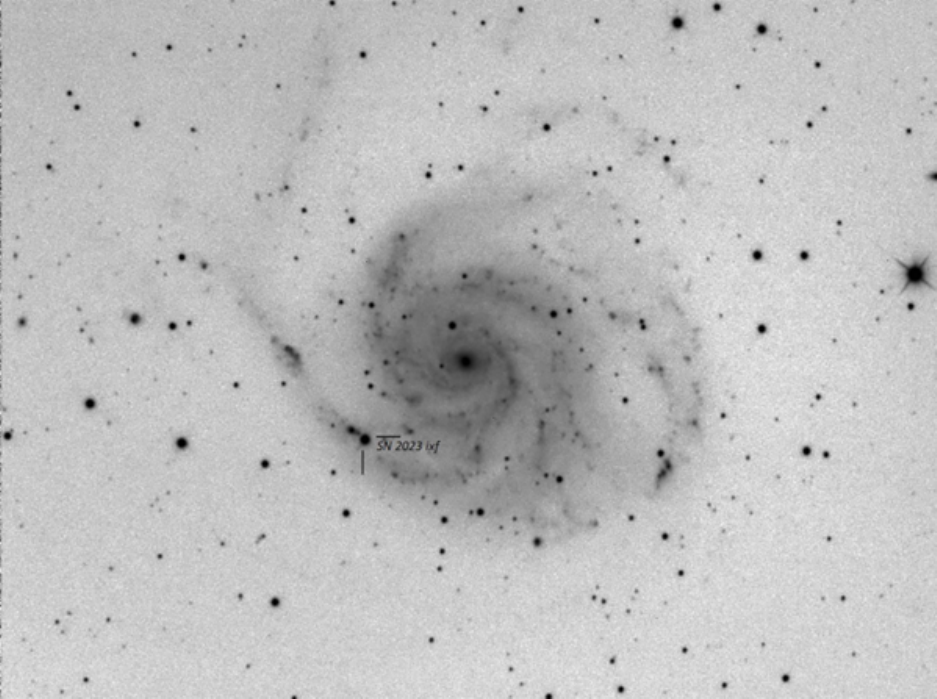}
\caption{This image at a higher resolution than Figure \ref{fig:fig1} made by one of us shows that SN 2023ixf occurred in a region rich in gas and near several star-forming regions. Image credit: Z. Organić, Orbanić Explorer Observatory.}
\label{fig:fig2}
\end{center}
\end{figure}

Ten different observatories were employed at different locations in Europe. Table \ref{set-up telescopi} (Appendix) shows the different types of telescopes and cameras used to make the measurements. In this preliminary study, the data collection did not apply any standard photometric conversion due to the large number of different filter sets used which included both standard and not standard sets (B (Johnson/Cousins), V (Johnson/Cousins), Rc (Johnson/Cousins), r Sloan, g Sloan, as well TriColor R, G, B, and Clear). Where possible, the exposure times were constrained so that the ADU values were approximately 50\% of the maximun ADU of the sensor, in order to obtain the highest sensibility and to be in the linear relation of the sensor itself. (Figure \ref{fig:fig1} and Figure \ref{fig:fig2}).

%\pagebreak

%\begin{figure}[!h]
% \includegraphics[width=\linewidth]{PLOT_VRB.pdf}
%\caption{Multiband observations of instrumental magnitude}
% \label{fig:plotvrb}
%\end{figure}

%\pagebreak

To obtain magnitudes, aperture photometry was performed by using MaximDL software \citep{george2000image}. The reference star was the star TYC 3852-1108-1 (V = 11.9, Type G1, \citep{gaia2016gaia}) for all observers. Observations made through the MicroObservatory in Arizona \citep{2023TNSAN.143....1F}, were also here presented. \\

Our photometric analysis resulted in a range of uncertainties of 0.002 to 0.06 magnitudes according to the different observatories, with a mean standard error equal to 0.02 - 0.007, according to the examined band magnitudes.

\section{Results}

\begin{figure}[ht!]
\begin{center}
\includegraphics[width=17cm]{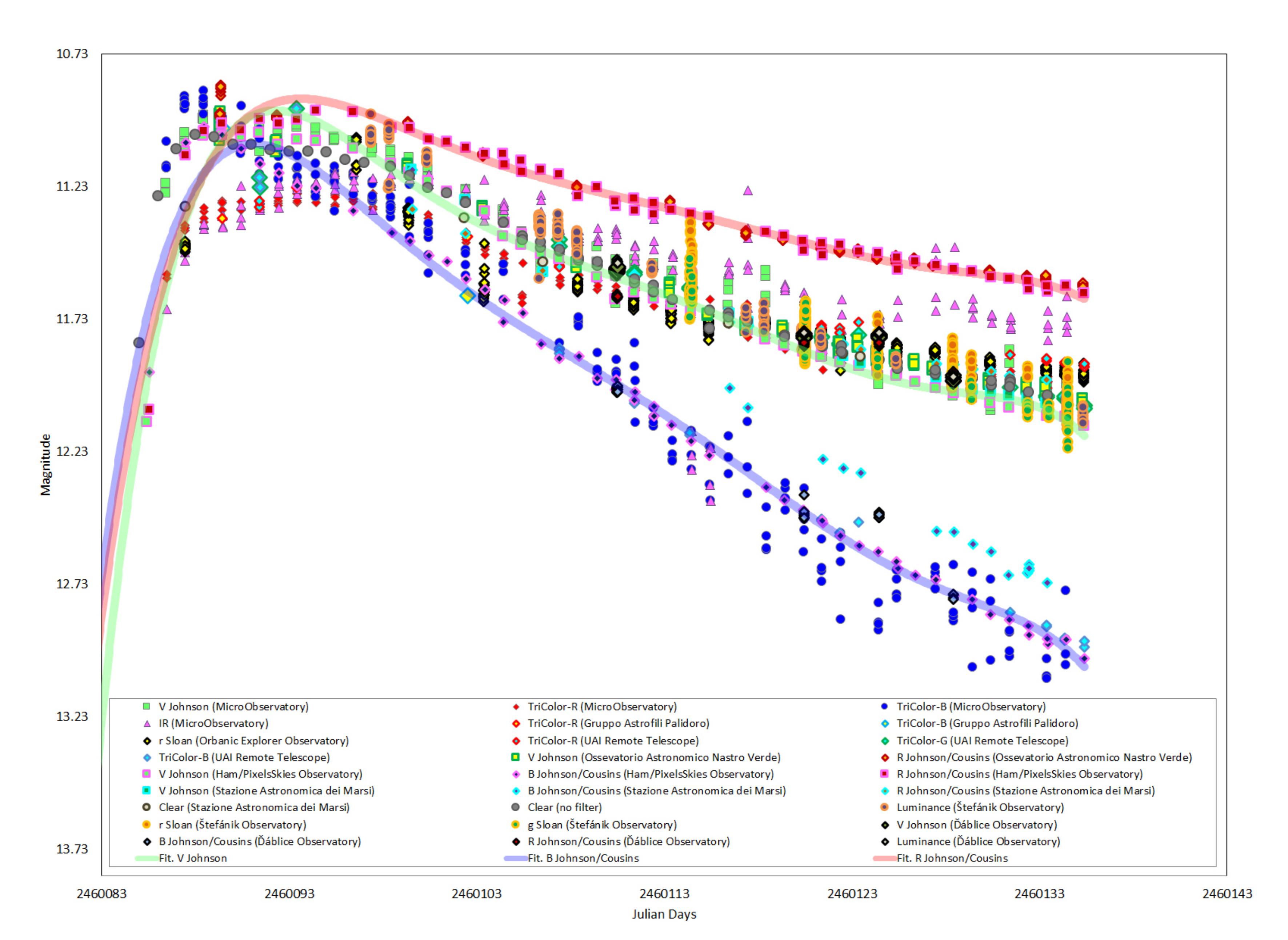}
\caption{The rise and the decline brightness decline of SN 2023ixf . The slope of the decrease in brightness of the observed SNe is different in each color band, becoming steadily shallower as we move to redder wavelengths. For clarity, error bars are not plotted }
\label{fig:plotvrb}
\end{center}
\end{figure}

Using the magnitude measurements performed from all the various observatories we can observe a rapid increase of brightness in the first 5 days after the discovery of the supernova, followed by a slow brightness decrease that appears clearly with a different slope for each filter (Figure \ref{fig:plotvrb}).

More precisely, Figure \ref{fig:plotvrb} shows the rapid ascent towards the maximum (2.5 magnitudes about in five days in the B filter). \\

The fading brightness seen in the B filter starts immediately after maximum with a linear slope of 0.0425 ± 0.02/day, while in the G filter the decrease started some days after the initial decline reported by B filter, with a slope equal to 0.022 ± 0.01/day.\\

The decline in the R filter began later still with a shallower slope equal to 0.0187 ± 0.007/day. Interestingly, 50 days after the explosion the decrease in magnitude seen in the B filter reached 2.5 magnitudes.
%The response difference by each filter applied to the observation of the supernova SN 2023ixf, led us to an analysis on the differences between photometric filters as “color indexes”. This study focused on the comparison of B-V and V-R indices, shown in Figure \ref{fig:plotindex}.

\pagebreak

\begin{figure}[ht!]
\begin{center}
\includegraphics[width=17cm]{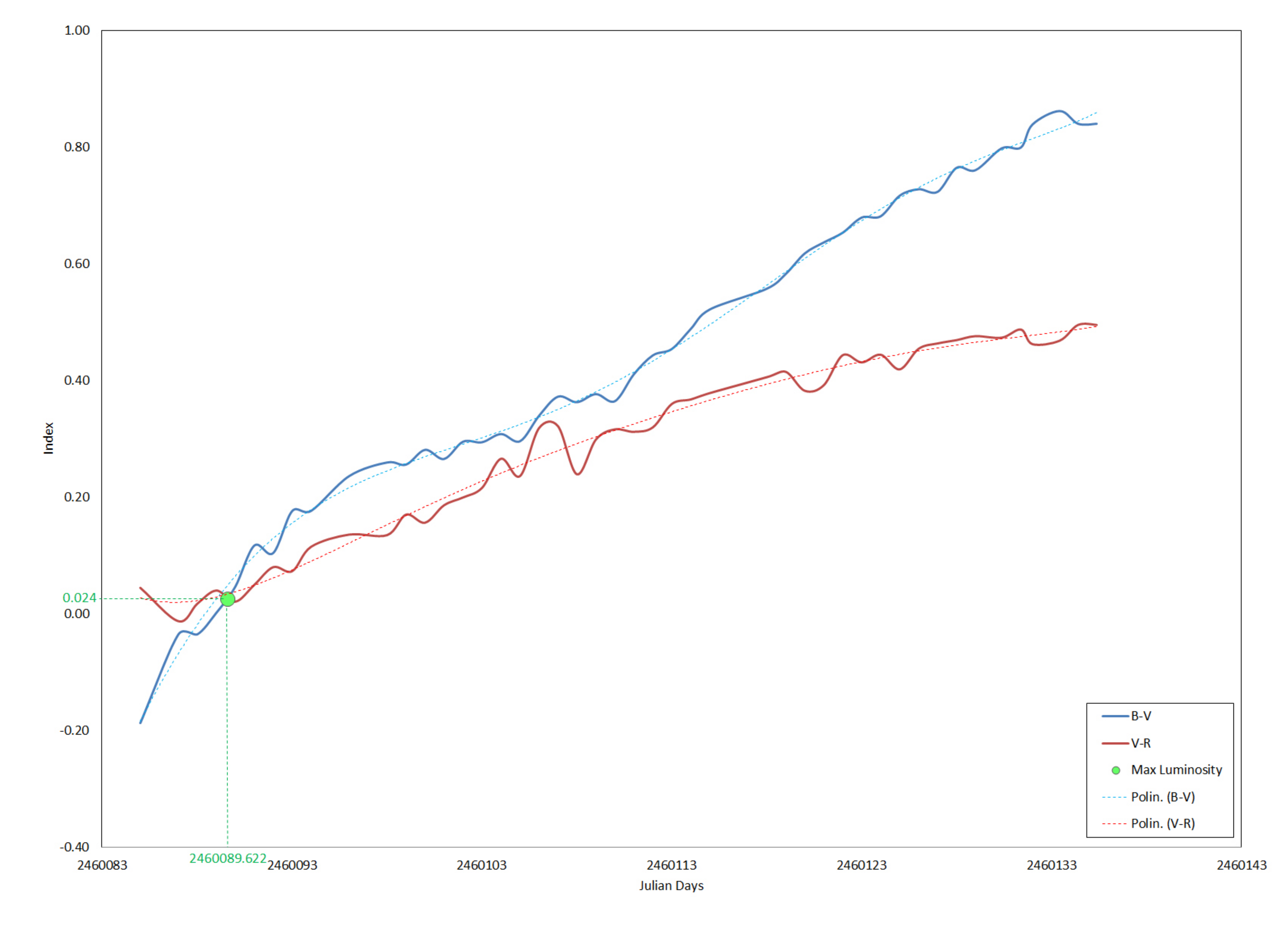}
\caption{Color indices B-V and V-R linear trend ($R^2 >$  0.95)}
\label{fig:plotindex}
\end{center}
\end{figure}

The response differences seen in each filter applied to the observation of the supernova SN 2023ixf, led to an analysis of the differences between photometric filters as `colour indexes”. This study focused on the comparison of B-V and V-R indices, shown in Figure \ref{fig:plotindex}.

%\pagebreak

In Figure \ref{fig:plotindex} clearly appears a huge variation of the B-V colour indices during the first 50 days after explosion from -0.20 ± 0.02 to +0.85 ± 0.02 and, similarly, V-R behaviour change from 0 ± 0.01 to +0.50 ± 0.01, presumably
linked to the cooling of the stellar photosphere.

Interestingly,the measurements show a point corresponding to May 25, 2023 ($JD = 2460089.622$) where B-V = V-R = 0.024. Next, we see an increasing trend as would be expected from a hot blackbody which is cooling, to the extent where most of the brightness in the B band has shifted towards the V and R bands \citep{1965ApJ...142.1317P}.

%\pagebreak

\hfill \break

\section{Discussion} \label{sec:discussion}

SN 2023ixf is a type II (core collapse) supernova located in the Pinwheel Galaxy (M101). It was first observed on May 19, 2023 by Koichi Itagaki and immediately classified as a type II supernova \citep{2023TNSAN.119....1P}.
The initial magnitude at discovery was 14.9. After discovery, the Zwicky Transient Facility project and observations from amateur astronomers found recovery images of the supernova, at magnitude 15.87 at less two days before discovery \citep{2023arXiv230606097H}. C.D. Kilpatric et al. \citep{2023arXiv230604722K} and J.E Jencson et al. \citep{2023arXiv230608678J} have proposed the identification of the progenitor, a variable Red Supergiant in the field, a study of the optical precursor that was recently updated by Y. Dong et al. \citep{2023arXiv230702539D}.
There are several sub-types within the Type II group: types II-L, II-P, II-b, and II-n. The L and P types refer to the shape of the light curve and how it fades (P stands for `plateau’, L for `linear’) \citep{2015A&A...582A...3G}. By photometric work, as seen in this paper, two principal divisions of Type II SNe can be made, the II-plateau (P), about 2/3 of the Type II SNe known, and II-linear (L) types. The former displays an extended period (typically $\sim 100d$) during which the (VRI-band) brightness remains constant, usually to within $\lesssim 0.5mag$; conversely, type II-L SNe decline linearly at a rate of a few hundredths of a magnitude per day in the same period of time. Our photometric measurements, reported here, started the same day of the discovery of the star, observing the quick rise to the maximum in 5 days and the slow decline in magnitudes of the star during the first 50 days post-explosion. At the 50th day after the explosion, the star continued the decrease in its brightness and its extent reached 2.5 magnitudes in B band from the maximum of brightness, so we can propose from our data a type II-L subtype for SN 2023ixf.
We are fully aware that the sample that we have here assembled is not yet photometrically homogenous as necessary, due to the different techniques and accuracies of the various used sets, and, moreover with standard and non-standard filters. However, the purpose of our measurements was to understand the magnitude differences using different filters and to monitor the advent or not of the plateau in the observed type II supernova \citep{2023arXiv230609311B}.

\pagebreak

\section{Conclusion}

Surely, SN 2023ixf will be studied for years to come, being able to offer an unprecedented opportunity to understand the physics of a type II supernova explosion, if only for its high brightness. 

%\pagebreak

\hfill \break

\section{Appendix}

\begin{table}[ht]
\centering
\renewcommand\arraystretch{1.1}
\caption{Setup used to the measurements by Observatories/Observers}
\label{set-up telescopi}
\begin{tabular}{llccl}\hline
\bf Observatory & \bf Telescope & \bf Camera & \bf Filters & \bf Place \\
\hline \hline
UAI Remote Telescope		          &Newton 0.20 m F/4 				 & Sbig ST10XME			& TriColor-R, TriColor-G,		    & Castiglione\\
					  & 				 &			& TriColor-B		    &  Del Lago Italy\\
\hline
Stazione Astronomica                      &0.36 m F/5.7                      &Sbig ST10XME           & Johnson/Cousins  &S.Jona\\
 dei Marsi					  &Ritchey-Chrétien			 & 			 & B, V, Rc          &    Italy\\
\hline
Gruppo Astrofili Palidoro                 &Newton 0.13 m F/5		 &           Canon 50D   &	TriColor-R  &Cerveteri\\
 					  &				 &                       &	 TriColor-G, TriColor-B &Italy\\
\hline
Orbanić Explorer - MPC                 &  Newton Truss                & Atik 414 Ex Mono      & r Sloan                 &    Pula   \\
code M19 				  & 0.25 m F/3.8 		 & 			 & &Croatia \\

\hline
Osservatorio Nastro                       &  S-C Meade LX200 14"      &  Sbig ST10XME         & R Johnson/Cousins &   Sorrento \\
Verde - MPC code C82 			  &F/10 with focal reducer  & 			 & V Johnson        &    Italy\\
\hline

Ham Observatory				  & Celestron C9.25 EdgeHD 				 & SX-TRIUS 694 PRO			& V Johnson		    & Chichester\\
	                                  & with focal reducer				 &			& B ,R Johnson/Cousins		    & UK\\
\hline
Štefánik Observatory			  &Celestron RASA 11"            &ZWO ASI 533 color     &g Sloan, r Sloan& Prague \\
                                          &                              &        &     	L (UV-IR bloc)   &Czech Republic\\
\hline
Ďáblice Observatory                       &Newton 0.30 m                    &Moravian instruments  &B ,R Johnson/Cousins &Prague\\
                                          &                              &C3-26000 mono         &V Johnson, L    &Cz. Republic\\
\hline
 MicroObservatory                         &6" telescope              &CCD detector          &V, TriColor-R, &Amado \\
					  &				 &			&TriColor-B & Arizona\\
\hline
 PixelSkies                         &Celestron C11 EdgeHD              &SX-TRIUS 694 PRO          &V Johnson, &Castilléjar\\
					  &				 &			&B ,R Johnson/Cousins & Spain\\
\hline
\hline
\end{tabular}
\end{table}

%\pagebreak

%\begin{figure}[ht]
%\begin{center}
%\includegraphics[width=17cm]{M101.pdf}
%\caption{Supernova star field (reverse mode) in M101 galaxy from Orbanić Explorer Observatory}
%\label{fig:M101}
%\end{center}
%\end{figure}

\section{Acknowledgment}

We would like to express our gratitude to all employees/collaborators of both Prague observatories, who have been most helpful by assisting the observations.
We would like express our gratitude to Mary Dussault, who granted us the use of the images from MicroObservatory Robotic Telescope Network.

\hfill \break

\bibliography{sample63}{}
\bibliographystyle{aasjournal}



\end{document}